\documentclass[a4paper,11pt,titlepage,oneside]{book} 
\pdfoutput=1

\usepackage[T1]{fontenc}
\usepackage[utf8]{inputenc}

\usepackage{ae}               
\usepackage[pdftex]{graphicx}
\usepackage{vmargin}          
\usepackage{fancyhdr}         
\usepackage[font=footnotesize]{subfig}
\usepackage[hyphens]{url}
\usepackage[absolute,overlay]{textpos}
\usepackage{tikz}
\usepackage[english]{babel}
\usepackage{color}
\usepackage{longtable}
\usepackage{cite}

\usepackage[bottom]{footmisc}
\usepackage[section,subsection,subsubsection]{extraplaceins}
\usepackage{multirow}
\usepackage{array}
\usepackage{nomencl}
\usepackage{color}
\usepackage{xcolor}
\usepackage{setspace}

\makenomenclature

\makeglossary

\definecolor{darkred}{rgb}{0.5,0,0}
\definecolor{darkgreen}{rgb}{0,0.5,0}
\definecolor{darkblue}{rgb}{0,0,0.5}

\setmarginsrb
{2.5cm}   
{1.5cm}   
{2.5cm}   
{2cm}   
{20pt}    
{0.25in}  
{9pt}     
{0.3in}

\usepackage{chngcntr}
\counterwithin{figure}{section}
\counterwithin{table}{section}

\usepackage{authblk}

\newcommand\mytitle{On Benchmarking Intrusion Detection Systems in Virtualized Environments}
\title{Cloud Usage Patterns}

\newcommand\TRnumber{Technical Report: SPEC-RG-2013-002\\Version: 1.0}
\newcommand\WGname{SPEC RG IDS Benchmarking Working Group}

\newcommand\TRdate{June 26, 2013}
\newcommand\TRcentralURL{research.spec.org}
\newcommand\TRrightURL{www.spec.org}

\newcommand\numAuthors{5} 
\makeatletter
\newcommand\defcase[1]{\@namedef{mycase@\the\numexpr#1\relax}}
\newcommand\putAuthors[1]{\@nameuse{mycase@\the\numexpr#1\relax}}
\makeatother

\defcase{0}{
	\begin{textblock}{8}[0,0](\colsinglecentralX,\rowoneY)
		\centering
		\small{\textbf{No Author}}\\
	\end{textblock}	
}

\defcase{1}{
	\begin{textblock}{8}[0,0](\colsinglecentralX,\rowoneY)
		\centering
		\small{\textbf{\authorOneName}}\\
		\footnotesize
		\authorOneAffil
	\end{textblock}	
}
\defcase{2}{
	\begin{textblock}{\authorCellWidth}[0,0](\colDoubleLeftX,\rowoneY)
		\centering
		\small{\textbf{\authorOneName}}\\
		\footnotesize
		\authorOneAffil
	\end{textblock}	
	\begin{textblock}{\authorCellWidth}[0,0](\colDoubleRightX,\rowoneY)
		\centering
		\small{\textbf{\authorTwoName}}\\
		\footnotesize
		\authorTwoAffil
	\end{textblock}	
}
\defcase{3}{
	\begin{textblock}{\authorCellWidth}[0,0](\coloneX,\rowoneY)
		\centering
		\small{\textbf{\authorOneName}}\\
		\footnotesize
		\authorOneAffil
	\end{textblock}	
	\begin{textblock}{\authorCellWidth}[0,0](\coltwoX,\rowoneY)
		\centering
		\small{\textbf{\authorTwoName}}\\
		\footnotesize
		\authorTwoAffil
	\end{textblock}	
	\begin{textblock}{\authorCellWidth}[0,0](\colthreeX,\rowoneY)
		\centering
		\small{\textbf{\authorThreeName}}\\
		\footnotesize
		\authorThreeAffil
	\end{textblock}	
}
\defcase{4}{
	\begin{textblock}{\authorCellWidth}[0,0](\coloneX,\rowoneY)
		\centering
		\small{\textbf{\authorOneName}}\\
		\footnotesize
		\authorOneAffil
	\end{textblock}	
	\begin{textblock}{\authorCellWidth}[0,0](\coltwoX,\rowoneY)
		\centering
		\small{\textbf{\authorTwoName}}\\
		\footnotesize
		\authorTwoAffil
	\end{textblock}	
	\begin{textblock}{\authorCellWidth}[0,0](\colthreeX,\rowoneY)
		\centering
		\small{\textbf{\authorThreeName}}\\
		\footnotesize
		\authorThreeAffil
	\end{textblock}
	\begin{textblock}{8}[0,0](\colsinglecentralX,\rowtwoY)
		\centering
		\small{\textbf{\authorFourName}}\\
		\footnotesize
		\authorFourAffil
	\end{textblock}	
}
\defcase{5}{
	\begin{textblock}{\authorCellWidth}[0,0](\coloneX,\rowoneY)
		\centering
		\small{\textbf{\authorOneName}}\\
		\footnotesize
		\authorOneAffil
	\end{textblock}	
	\begin{textblock}{\authorCellWidth}[0,0](\coltwoX,\rowoneY)
		\centering
		\small{\textbf{\authorTwoName}}\\
		\footnotesize
		\authorTwoAffil
	\end{textblock}	
	\begin{textblock}{\authorCellWidth}[0,0](\colthreeX,\rowoneY)
		\centering
		\small{\textbf{\authorThreeName}}\\
		\footnotesize
		\authorThreeAffil
	\end{textblock}
	\begin{textblock}{\authorCellWidth}[0,0](\colDoubleLeftX,\rowtwoY)
		\centering
		\small{\textbf{\authorFourName}}\\
		\footnotesize
		\authorFourAffil
	\end{textblock}	
	\begin{textblock}{\authorCellWidth}[0,0](\colDoubleRightX,\rowtwoY)
		\centering
		\small{\textbf{\authorFiveName}}\\
		\footnotesize
		\authorFiveAffil
	\end{textblock}	
}
\defcase{6}{
	\begin{textblock}{\authorCellWidth}[0,0](\coloneX,\rowoneY)
		\centering
		\small{\textbf{\authorOneName}}\\
		\footnotesize
		\authorOneAffil
	\end{textblock}	
	\begin{textblock}{\authorCellWidth}[0,0](\coltwoX,\rowoneY)
		\centering
		\small{\textbf{\authorTwoName}}\\
		\footnotesize
		\authorTwoAffil
	\end{textblock}	
	\begin{textblock}{\authorCellWidth}[0,0](\colthreeX,\rowoneY)
		\centering
		\small{\textbf{\authorThreeName}}\\
		\footnotesize
		\authorThreeAffil
	\end{textblock}
	\begin{textblock}{\authorCellWidth}[0,0](\coloneX,\rowtwoY)
		\centering
		\small{\textbf{\authorFourName}}\\
		\footnotesize
		\authorFourAffil
	\end{textblock}	
	\begin{textblock}{\authorCellWidth}[0,0](\coltwoX,\rowtwoY)
		\centering
		\small{\textbf{\authorFiveName}}\\
		\footnotesize
		\authorFiveAffil
	\end{textblock}	
	\begin{textblock}{\authorCellWidth}[0,0](\colthreeX,\rowtwoY)
		\centering
		\small{\textbf{\authorSixName}}\\
		\footnotesize
		\authorSixAffil
	\end{textblock}
}
\defcase{7}{
	\begin{textblock}{\authorCellWidth}[0,0](\coloneX,\rowoneY)
		\centering
		\small{\textbf{\authorOneName}}\\
		\footnotesize
		\authorOneAffil
	\end{textblock}	
	\begin{textblock}{\authorCellWidth}[0,0](\coltwoX,\rowoneY)
		\centering
		\small{\textbf{\authorTwoName}}\\
		\footnotesize
		\authorTwoAffil
	\end{textblock}	
	\begin{textblock}{\authorCellWidth}[0,0](\colthreeX,\rowoneY)
		\centering
		\small{\textbf{\authorThreeName}}\\
		\footnotesize
		\authorThreeAffil
	\end{textblock}
	\begin{textblock}{\authorCellWidth}[0,0](\coloneX,\rowtwoY)
		\centering
		\small{\textbf{\authorFourName}}\\
		\footnotesize
		\authorFourAffil
	\end{textblock}	
	\begin{textblock}{\authorCellWidth}[0,0](\coltwoX,\rowtwoY)
		\centering
		\small{\textbf{\authorFiveName}}\\
		\footnotesize
		\authorFiveAffil
	\end{textblock}	
	\begin{textblock}{\authorCellWidth}[0,0](\colthreeX,\rowtwoY)
		\centering
		\small{\textbf{\authorSixName}}\\
		\footnotesize
		\authorSixAffil
	\end{textblock}
	\begin{textblock}{8}[0,0](\colsinglecentralX,\rowthreeY)
		\centering
		\small{\textbf{\authorSevenName}}\\
		\footnotesize
		\authorSevenAffil
	\end{textblock}
}
\defcase{8}{
	\begin{textblock}{\authorCellWidth}[0,0](\coloneX,\rowoneY)
		\centering
		\small{\textbf{\authorOneName}}\\
		\footnotesize
		\authorOneAffil
	\end{textblock}	
	\begin{textblock}{\authorCellWidth}[0,0](\coltwoX,\rowoneY)
		\centering
		\small{\textbf{\authorTwoName}}\\
		\footnotesize
		\authorTwoAffil
	\end{textblock}	
	\begin{textblock}{\authorCellWidth}[0,0](\colthreeX,\rowoneY)
		\centering
		\small{\textbf{\authorThreeName}}\\
		\footnotesize
		\authorThreeAffil
	\end{textblock}
	\begin{textblock}{\authorCellWidth}[0,0](\coloneX,\rowtwoY)
		\centering
		\small{\textbf{\authorFourName}}\\
		\footnotesize
		\authorFourAffil
	\end{textblock}	
	\begin{textblock}{\authorCellWidth}[0,0](\coltwoX,\rowtwoY)
		\centering
		\small{\textbf{\authorFiveName}}\\
		\footnotesize
		\authorFiveAffil
	\end{textblock}	
	\begin{textblock}{\authorCellWidth}[0,0](\colthreeX,\rowtwoY)
		\centering
		\small{\textbf{\authorSixName}}\\
		\footnotesize
		\authorSixAffil
	\end{textblock}
	\begin{textblock}{\authorCellWidth}[0,0](\colDoubleLeftX,\rowthreeY)
		\centering
		\small{\textbf{\authorSevenName}}\\
		\footnotesize
		\authorSevenAffil
	\end{textblock}
	\begin{textblock}{\authorCellWidth}[0,0](\colDoubleRightX,\rowthreeY)
		\centering
		\small{\textbf{\authorEightName}}\\
		\footnotesize
		\authorEightAffil
	\end{textblock}
}
\defcase{9}{
	\begin{textblock}{\authorCellWidth}[0,0](\coloneX,\rowoneY)
		\centering
		\small{\textbf{\authorOneName}}\\
		\footnotesize
		\authorOneAffil
	\end{textblock}	
	\begin{textblock}{\authorCellWidth}[0,0](\coltwoX,\rowoneY)
		\centering
		\small{\textbf{\authorTwoName}}\\
		\footnotesize
		\authorTwoAffil
	\end{textblock}	
	\begin{textblock}{\authorCellWidth}[0,0](\colthreeX,\rowoneY)
		\centering
		\small{\textbf{\authorThreeName}}\\
		\footnotesize
		\authorThreeAffil
	\end{textblock}
	\begin{textblock}{\authorCellWidth}[0,0](\coloneX,\rowtwoY)
		\centering
		\small{\textbf{\authorFourName}}\\
		\footnotesize
		\authorFourAffil
	\end{textblock}	
	\begin{textblock}{\authorCellWidth}[0,0](\coltwoX,\rowtwoY)
		\centering
		\small{\textbf{\authorFiveName}}\\
		\footnotesize
		\authorFiveAffil
	\end{textblock}	
	\begin{textblock}{\authorCellWidth}[0,0](\colthreeX,\rowtwoY)
		\centering
		\small{\textbf{\authorSixName}}\\
		\footnotesize
		\authorSixAffil
	\end{textblock}
	\begin{textblock}{\authorCellWidth}[0,0](\coloneX,\rowthreeY)
		\centering
		\small{\textbf{\authorSevenName}}\\
		\footnotesize
		\authorSevenAffil
	\end{textblock}	
	\begin{textblock}{\authorCellWidth}[0,0](\coltwoX,\rowthreeY)
		\centering
		\small{\textbf{\authorEightName}}\\
		\footnotesize
		\authorEightAffil
	\end{textblock}	
	\begin{textblock}{\authorCellWidth}[0,0](\colthreeX,\rowthreeY)
		\centering
		\small{\textbf{\authorNineName}}\\
		\footnotesize
		\authorNineAffil
	\end{textblock}
}
\defcase{10}{
	\begin{textblock}{\authorCellWidth}[0,0](\coloneX,\rowoneY)
		\centering
		\small{\textbf{\authorOneName}}\\
		\footnotesize
		\authorOneAffil
	\end{textblock}	
	\begin{textblock}{\authorCellWidth}[0,0](\coltwoX,\rowoneY)
		\centering
		\small{\textbf{\authorTwoName}}\\
		\footnotesize
		\authorTwoAffil
	\end{textblock}	
	\begin{textblock}{\authorCellWidth}[0,0](\colthreeX,\rowoneY)
		\centering
		\small{\textbf{\authorThreeName}}\\
		\footnotesize
		\authorThreeAffil
	\end{textblock}
	\begin{textblock}{\authorCellWidth}[0,0](\coloneX,\rowtwoY)
		\centering
		\small{\textbf{\authorFourName}}\\
		\footnotesize
		\authorFourAffil
	\end{textblock}	
	\begin{textblock}{\authorCellWidth}[0,0](\coltwoX,\rowtwoY)
		\centering
		\small{\textbf{\authorFiveName}}\\
		\footnotesize
		\authorFiveAffil
	\end{textblock}	
	\begin{textblock}{\authorCellWidth}[0,0](\colthreeX,\rowtwoY)
		\centering
		\small{\textbf{\authorSixName}}\\
		\footnotesize
		\authorSixAffil
	\end{textblock}
	\begin{textblock}{\authorCellWidth}[0,0](\coloneX,\rowthreeY)
		\centering
		\small{\textbf{\authorSevenName}}\\
		\footnotesize
		\authorSevenAffil
	\end{textblock}	
	\begin{textblock}{\authorCellWidth}[0,0](\coltwoX,\rowthreeY)
		\centering
		\small{\textbf{\authorEightName}}\\
		\footnotesize
		\authorEightAffil
	\end{textblock}	
	\begin{textblock}{\authorCellWidth}[0,0](\colthreeX,\rowthreeY)
		\centering
		\small{\textbf{\authorNineName}}\\
		\footnotesize
		\authorNineAffil
	\end{textblock}
	\begin{textblock}{8}[0,0](\colsinglecentralX,\rowfourY)
		\centering
		\small{\textbf{\authorTenName}}\\
		\footnotesize
		\authorTenAffil
	\end{textblock}
}

\newcommand\authorOneName{Aleksandar Milenkoski}
\newcommand\authorOneAffil{
	Institute for Program Structures and Data Organization\\
    Karlsruhe Institute of Technology\\
    Karlsruhe, Germany\\
    \emph{milenkoski@kit.edu}\\
}

\newcommand\authorTwoName{Samuel Kounev}
\newcommand\authorTwoAffil{
	Institute for Program Structures and Data Organization\\
    Karlsruhe Institute of Technology\\
    Karlsruhe, Germany\\
    \emph{kounev@kit.edu}\\
}

\newcommand\authorThreeName{Alberto Avritzer}
\newcommand\authorThreeAffil{
	Siemens Corporate Research\\
	Princeton, NJ USA\\
	\emph{alberto.avritzer@siemens.com}\\
}

\newcommand\authorFourName{Nuno Antunes}
\newcommand\authorFourAffil{
	CISUC, Department of Informatics Engineering\\
	University of Coimbra\\
	Coimbra, Portugal\\
	\emph{nmsa@dei.uc.pt}
}

\newcommand\authorFiveName{Marco Vieira}
\newcommand\authorFiveAffil{
	CISUC, Department of Informatics Engineering\\
	University of Coimbra\\
	Coimbra, Portugal\\
	\emph{mvieira@dei.uc.pt}
}

\newcommand\authorSixName{FirstName6 LastName6}
\newcommand\authorSixAffil{
	Department of Cloud Computing3,\\
	State University of SomeCity3,\\
	SomeCity'sLongerName, LongLongLongCountryName4,\\
	e-mail2@e-mail.com
}
\newcommand\authorSevenName{John Doe}
\newcommand\authorSevenAffil{
	Institute of Informatics,\\
	University of Polar Cirlce,\\
	Acity, Acoutry,\\
	john@upc.edu
}
\newcommand\authorEightName{FirstName8 LastName8}
\newcommand\authorEightAffil{
	Department of Cloud Computing3,\\
	State University of SomeCity3,\\
	SomeCity'sLongerName, LongLongLongCountryName4,\\
	e-mail2@e-mail.com
}
\newcommand\authorNineName{FirstName9 LastName9}
\newcommand\authorNineAffil{
	Department of Cloud Computing3,\\
	State University of SomeCity3,\\
	SomeCity'sLongerName, LongLongLongCountryName4,\\
	e-mail2@e-mail.com
}
\newcommand\authorTenName{FirstName10 LastName10}
\newcommand\authorTenAffil{
	Department of Cloud Computing3,\\
	State University of SomeCity3,\\
	SomeCity'sLongerName, LongLongLongCountryName4,\\
	e-mail2@e-mail.com
}

\begin{document}
 
\selectlanguage{english} 
\frontmatter

\thispagestyle{empty}
\newcommand{\changefont}[3]{\fontfamily{#1} \fontseries{#2} \fontshape{#3} \selectfont}
\newcommand{\diameter}{20}
\newcommand{\xone}{-25}
\newcommand{\xtwo}{165}
\newcommand{\yone}{20}
\newcommand{\ytwo}{-253}

\newcommand{\rowoneY}{5.5}		
\newcommand{\rowtwoY}{7.0}
\newcommand{\rowthreeY}{8.5}
\newcommand{\rowfourY}{10.1}

\newcommand{\coloneX}{2.5}
\newcommand{\coltwoX}{7.45}
\newcommand{\colthreeX}{12.4}

\newcommand{\colDoubleLeftX}{5}
\newcommand{\colDoubleRightX}{10}

\newcommand{\colsinglecentralX}{5.9}

\newcommand{\authorCellWidth}{4.9}

\begin{titlepage}
\begin{tikzpicture}[overlay]
\draw[color=gray]  
 (\xone mm, \yone mm) -- (\xtwo mm, \yone mm) arc (90:0:\diameter pt) 
  -- (\xtwo mm + \diameter pt , \ytwo mm) -- (\xone mm + \diameter pt , \ytwo mm) 
 arc (270:180:\diameter pt) -- (\xone mm, \yone mm);
\end{tikzpicture}

\changefont{phv}{m}{n}	
\begin{textblock}{14}[0,0](3,2.3)
	\centering
	\large{\TRnumber}\\
	\vspace*{1cm}
	\huge{\mytitle}\\
	\vspace*{0.5cm}
	\Large{\WGname}
\end{textblock}
\begin{textblock}{15.5}[0,0](2,5.2)
	\begin{tikzpicture}
		\fill[red!80!brown] (0,0cm) rectangle (19.5cm,0.1cm);
	\end{tikzpicture}
\end{textblock}

\begin{center}
	\putAuthors{\numAuthors}
\end{center}

\begin{textblock}{14}[0,0](3,13)
	\hfill
	\includegraphics[width=3cm]{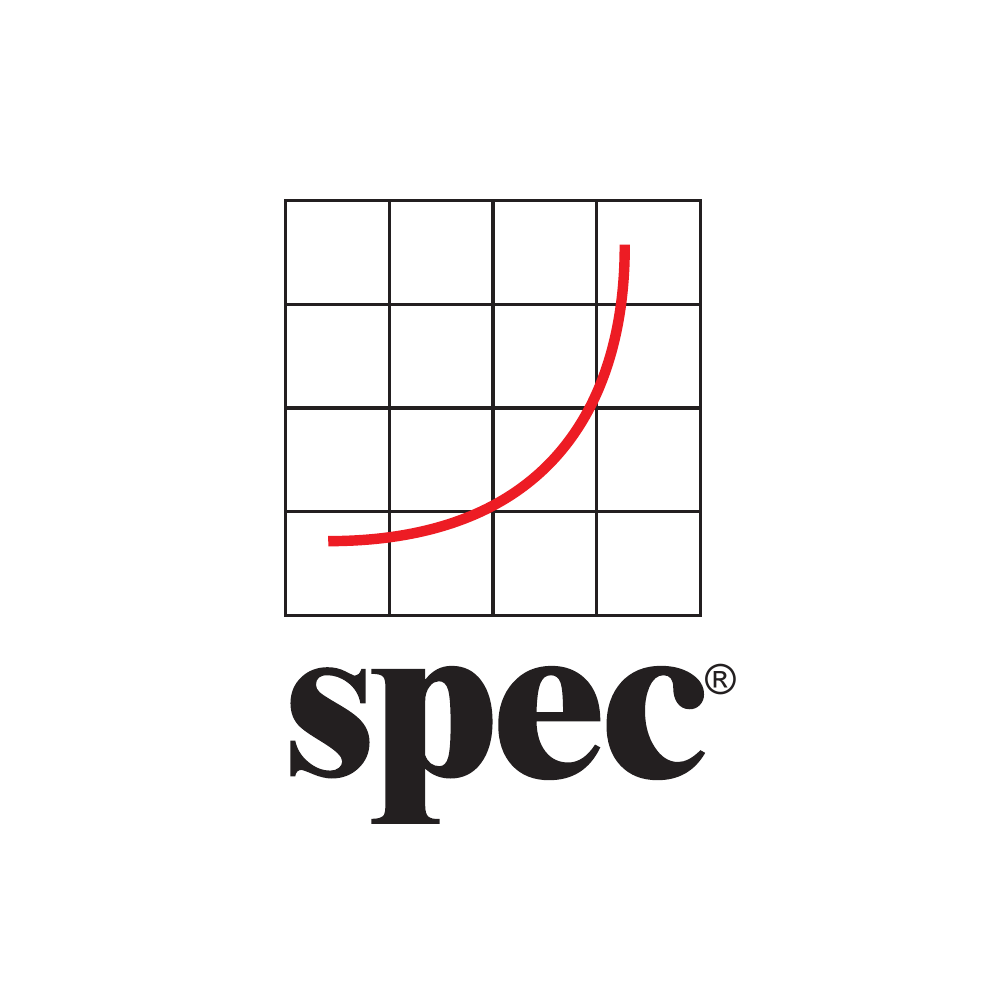} \hfill
	\includegraphics[width=1.9cm]{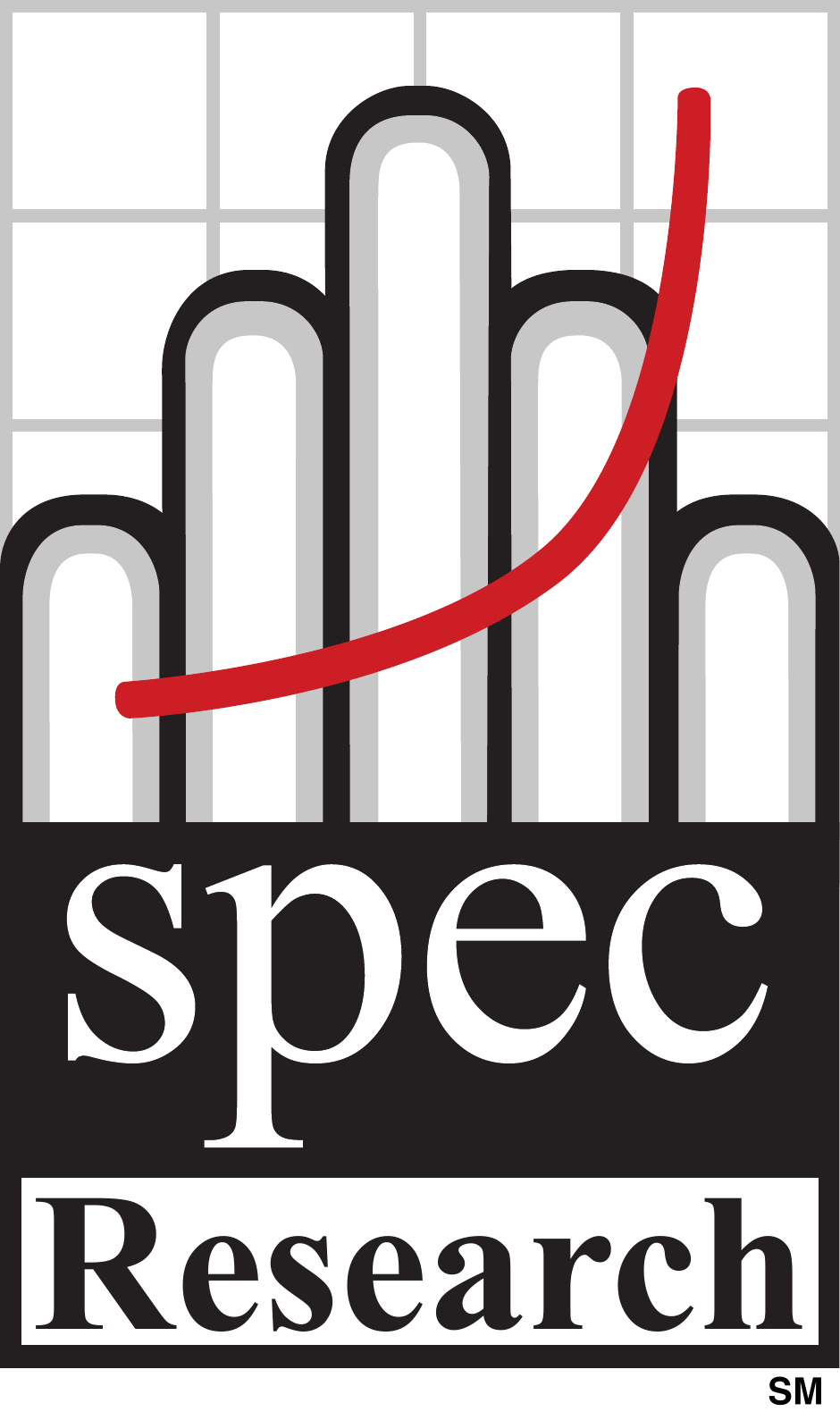} \hspace{1.1cm}\hfill
	\hfill
\end{textblock}

\begin{textblock}{14}[0,0](3,16.75)
	\centering
	\large{\textbf{\TRdate}}
	\hfill
	\large{\textbf{\TRcentralURL}}
	\hfill
	\large{\textbf{\TRrightURL}}
\end{textblock}

\end{titlepage}

\newpage 
\thispagestyle{empty}
\mbox{}

\newpage
\pagenumbering{roman}
\setcounter{tocdepth}{4}
\begin{spacing}{1.5}
 \tableofcontents
 \end{spacing}

\newpage
\thispagestyle{plain}
\section*{Executive Summary}

Modern intrusion detection systems (IDSes) for virtualized environments are deployed in the virtualization layer with components inside the virtual machine monitor (VMM) and the trusted host virtual machine (VM). Such IDSes can monitor at the same time the network and host activities of all guest VMs running on top of a VMM being isolated from malicious users of these VMs. We refer to IDSes for virtualized environments as \emph{VMM-based IDSes}. In this work, we analyze state-of-the-art intrusion detection techniques applied in virtualized environments and architectures of VMM-based IDSes. Further, we identify challenges that apply specifically to benchmarking VMM-based IDSes focussing on workloads and metrics. For example, we discuss the challenge of defining representative baseline benign workload profiles as well as the challenge of defining malicious workloads containing attacks targeted at the VMM. We also discuss the impact of on-demand resource provisioning features of virtualized environments (e.g., CPU and memory hotplugging, memory ballooning) on IDS benchmarking measures such as capacity and attack detection accuracy. Finally, we outline future research directions in the area of benchmarking VMM-based IDSes and of intrusion detection in virtualized environments in general.

\vspace{0.5cm}

\textbf{Keywords\footnote{The used keywords are defined as part of The 2012 ACM Computing Classification System \cite{acm:classification}.}:} \\
Security and Privacy - Intrusion/anomaly detection and malware mitigation  -  Intrusion detection systems\\
Security and Privacy - Systems security  -  Operating systems security - Virtualization and security\\
General and reference - Cross-computing tools and techniques - Evaluation\\
General and reference - Cross-computing tools and techniques - Metrics

\mainmatter
\pagenumbering{arabic}
\include{intro}

\section{Introduction}
\label{sec:introduction}

An intrusion detection system (IDS) increases the security of the environment in which it is deployed by enabling the detection of activities with malicious intent. The detection of such activities, referred to as attacks, enables responsive actions which could stop an on-going attack, or mitigate the damage of a successful attack. However, the effectiveness of an IDS is challenged by the innovative and continuously evolving attack techniques that target novel platforms and technologies. For instance, the cloud computing paradigm, which is based on virtualization as a key enabling technology, is rapidly gaining in popularity making cloud environments an attractive target for attackers.  The latter can exploit novel attack venues resulting from the use of virtualization or target vulnerabilities in virtualization platforms themselves. For instance, recent reports indicate the existence of high-impact attacks against virtual machine monitors (VMMs) that exploit specific operational or configuration weaknesses \cite{ibm:midterm}, \cite{informationweek:vulnerability}. In order to keep up with the ever-evolving security threat landscape, novel IDS architectures and intrusion detection techniques are needed. The research and industrial communities have developed innovative IDSes designed specifically to operate in virtualized environments. Such IDSes reside in the virtualization layer (i.e., they leverage the functionalities of the VMM) and thus, they are able to monitor multiple guest virtual machines (VMs) at the same time. In this work, we refer to such IDSes as \emph{VMM-based IDSes}.

When considering to deploy an IDS in a given target computing environment, one normally aims to deploy a carefully chosen IDS that operates optimally in the specific environment in order to reduce the chance of security breaches. Further, the efficiency of any deployed IDS is known to be extremely sensitive to the configuration of the IDS itself. Thus, a common goal is to identify an optimal IDS configuration. Benchmarking of IDSes contributes towards addressing these issues by enabling the comparison of multiple IDSes, or of multiple configurations of a single IDS, with respect to various IDS properties such as attack detection accuracy, resource consumption, and so on. Therefore, the research area of benchmarking IDSes has recieved a great deal of attention over the last decade. Researchers have developed many benchmarking methodologies \cite{hall:capacity}, metrics \cite{gu:measuring}, \cite{gaffney:evaluation}, and workload generation methods \cite{fonesca:vulnerability}, \cite{sommers:aframework} for use in IDS benchmarking tests.  However, many issues, such as the provisioning of representative malicious workloads and the generation of background benign workloads in a scientically rigorous manner, still persist on the IDS benchmarking scene \cite{mell:anoverview}, \cite{zanero:my}, \cite{stuckman:tracking}. Further, the novel architectures of modern IDSes for virtualized environments warrant novel benchmarking approaches that are riddled with many challenges. The lack of appropriate novel benchmarking methodologies, representative workloads, and representative metrics results in inability to accurately evaluate such IDSes in terms of their various security- and performance-related properties. 

In this work, we provide a survey of intrusion detection practices applied in virtualized environments. We analyze state-of-the-art IDSes and intrusion detection techniques employed in such environments. Further, we identify and discuss major challenges that apply specifically to benchmarking VMM-based IDSes, with a focus on workloads and metrics in particular. We also provide an outlook on future developments in the research area of benchmarking VMM-based IDSes and of intrusion detection in virtualized environments in general. 

This work is organized as follows: In Section~\ref{sec:intrusion_detection_in_virtualized_environments}, we analyze architectures of IDSes for virtualized environments and various intrusion detection techniques that are used in this context. In Section~\ref{sec:challenges_on_benchmarking}, we investigate the open challenges related to benchmarking such IDSes. Finally, in Section~\ref{sec:conclusion}, we present our conclusions.

\section{Intrusion Detection in Virtualized Environments}
\label{sec:intrusion_detection_in_virtualized_environments}

\subsection{VMM-Based Intrusion Detection Systems}
\label{sec:vmm_based}

Modern IDSes for virtualized environments leverage the functionalities of virtualization platforms, i.e., of the underlying VMMs, and are thus able to monitor multiple guest VMs at the same time. In a virtualized environment, the VMM is deployed between the guest VMs and the shared physical hardware, and therefore, it plays the role of an intermediary that enables the operation of the guest VMs. Thus, a VMM is a suitable location for the deployment of a monitoring agent that would monitor both the host and the network activities of guest VMs. Also, the analysis and the control module of a typical VMM-based IDS are normally deployed in the administrative domain of the virtualized environment. Given that VMM-based IDSes perform many intrusion detection activities (e.g., monitoring, analysis) in the virtualization layer instead of directly in the monitored guest VMs, they possess two features that distinguish them from the IDSes for traditional environments: isolation from, and transparency to, attackers. The isolation of a VMM-based IDS from attackers protect it from direct attacks against it that might result in crashing or disabling the IDS itself. Further, since VMM-based IDSes are able to perform host and network intrusion detection without being directly deployed in the guest VMs, they are transparent, i.e., hardly detectable by attackers\footnote{Probing attacks for detection of VMM-based IDSes have recently started to emerge. One such an attack is consisting of measuring the execution time of system calls in a guest VM, where an execution delay indicates IDS presence. However, this and similar IDS probing attacks can be easily prevented, for example, the ACPS IDS uses the SWADR (synchronous warning - asynchronous detection and response) approach \cite{lombardi:secure}. Therefore, although a certain risk of detection exists, VMM-based IDSes are still considered as transparent when compared to IDSes for traditional environments.}.  This additionally decreases the possibility for a successful attack against the IDS itself. To the contrary, for example, host-based IDSes for traditional environments usually operate as active processes in the same host operating systems that they monitor, and therefore, their presence can often be detected by attackers in a straightforward manner. 

However, VMM-based IDSes face a major challenge that they need to handle in order to be effective. That is, the virtualization layer provides only low-level hardware information about the guest VMs (e.g., CPU register values, memory content), as opposed to high-level information at the OS-level needed as input to the intrusion detection logic (e.g., executed system calls, active processes, filesystem alternations, and similar). This problem is known as the \emph{semantic gap} problem and is a subject of extensive research\footnote{Note that the semantic gap issue relates mainly to IDS input data for host intrusion detection, a feature of many existing VMM-based IDSes. To the contrary, for network intrusion detection, VMM-based IDSes normally tap into a network interface card used by multiple guest VMs, thus monitoring the network activities of all guest VMs at the same time.}. To tackle the semantic gap problem, two different architectures of VMM-based IDSes exist: (i) \emph{intrusive} IDSes (i.e., IDSes that have monitoring components, commonly referred to as ``hooks'', deployed inside guest VMs), and (ii) \emph{non-intrusive} IDSes (i.e., IDSes that do not use monitoring components deployed inside guest VMs).

\begin{figure*}[th]

\subfloat[]{\includegraphics[scale=0.92]{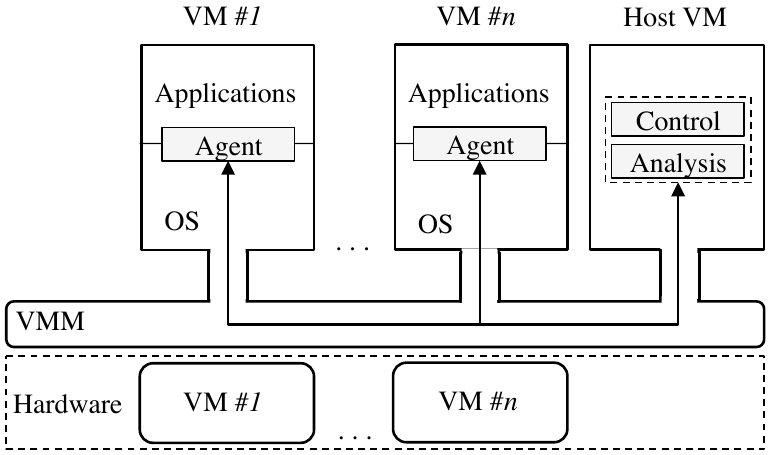}
\label{fig:intrusive}}
\hfil
\subfloat[]{\includegraphics[scale=0.92]{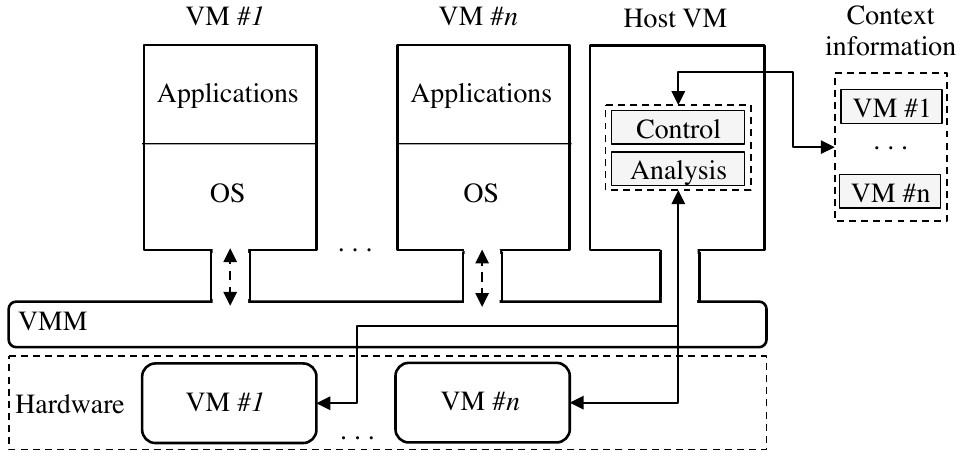}
\label{fig:non_intrusive}}
\caption{Architecture of a) intrusive, and b) non-intrusive VMM-based IDS.}
\label{fig:recording}
\end{figure*}

The \textbf{intrusive} VMM-based IDSes have monitoring agents deployed inside guest VMs in order to directly access OS-level information about the monitored systems. The agents normally deliver such information to the analysis module of a VMM-based IDS, where the intrusion logic is executed. Although the deployment of agents inside the guest VMs is an intuitive and straightforward approach towards closing the semantic gap, it has the disadvantage of exposing IDS components to malicious users of guest VMs; that is, an intrusive VMM-based IDS achieves only partial isolation and transparency making IDS detection and subversion more easy. On the other hand, the deployment of agents inside guest VMs provides access to a large amount of rich OS-level information enabling efficient intrusion detection. In Figure~\ref{fig:intrusive}, we depict an architecture of a typical intrusive VMM-based IDS. The depicted IDS architecture consists of user-space guest VM agents that monitor application activities directly without accessing the hardware resources allocated to each VM. In the following, we show that the need of accessing hardware resources allocated to guest VMs is the major difference between intrusive and non-intrusive approaches. Some existing intrusive VMM-based IDSes are described in Asrigo et al. \cite{asrigo:using} and Payne et al. \cite{payne:lares}.

\textbf{Non-intrusive} IDSes, in contrast to intrusive VMM-based IDSes, use \emph{virtual machine introspection} (VMI) in order to obtain information about guest VMs without using monitoring agents inside guest VMs. A typical VMI procedure normally consists of obtaining two types of information: (i) hardware-level information for each VM, which can be obtained at the virtualization platform level, and (ii) high-level, often OS-specific, context information (e.g., filesystem structure, memory regions that store information on active processess, and similar). The context information enables interpretation of the hardware-level information, thus making it useful for intrusion detection (e.g., in order for a VMM-based IDS to obtain file modification dates, knowledge about the filesystem used by a guest VM may help in parsing relevant file information stored in specific memory regions). VMM-based IDSes normally use VMI libraries and/or tools built for a specific virtualization platform (e.g., VMware, Xen). Some of these libraries and tools obtain only low-level hardware information, while others also obtain context information about guest VMs. For instance, AntFarm \cite{jones:antfarm} is a VMI framework that provides OS process information and XenAccess \cite{xenaccess} is a VMI library for Xen that provides hardware-level information (e.g., CPU register values, memory content). Note that when a VMM-based IDS uses a VMI library/tool providing only hardware-level information, context information is normally either hard-coded into the IDS itself or specified as part of the configuration space of the IDS. We depict an architecture of a typical non-intrusive VMM-based IDS in Figure~\ref{fig:non_intrusive}. 

The VMI procedures normally exhibit high complexity due to the challenge of reconstructing high-level data relevant for intrusion detection based on hardware-level and OS-level context information. Dehnert \cite{dehnert:intrusion}, when describing a non-intrusive VMM-based IDS leveraging VMware's VProbes technology \cite{vprobes}, states: ``\textit{The first step in implementing the gatherer [a monitoring component] is to find where the Linux or Microsoft Windows kernel stores the pertinent data. While a userspace IDS could use relatively well-defined, clearly documented, and stable interfaces such as system calls, or read /proc to gather the required information, we are unable to run code from the target [monitored] system. As a result, we must directly access kernel memory. Determining the relevant structures is a process that involves reading the source, disassembling system call implementations, or looking at debugging symbols.}''. Next, we briefly describe the operation of Wizard \cite{srivastava:secure}, a non-intrusive VMM-based IDS monitoring the execution of system calls in order to detect attacks that alter the kernel behavior. 

Since system calls are executed at the OS-level while Wizard operates at the virtualization level, Wizard intercepts the VM calls generated by kernel service handlers of guest VMs. Wizard maps VM calls to system calls by leveraging stored information on such a mapping obtained during a training process that requires OS-specific knowledge about the guest VMs in a given virtualized environment. Quoting from Srivastava et al. \cite{srivastava:secure}: ``\textit{We need to know only the guest OS' software interrupt number and the hardware register it uses to store the specific service requested by software.}'' Each time Wizard intercepts a VM call, it obtains hardware-level information; that is, Wizard reads the value stored in the CR3 register in order to map such a call to a specific guest VM process\footnote{The CR3 control register on Intel x86 platforms stores the page table base address, which is unique for each process enabling the identification of the process that executes a given system call.}.

\subsection{Intrusion Detection Techniques}
\label{sec:intrusion_detection_techniques}

As it is common in the field, we distinguish between misuse-based and anomaly-based intrusion detection techniques. In this section, we discuss the use of these techniques in virtualized environments in terms of features and benefits that they offer. 

\subsubsection{Misuse-based Intrusion Detection}
\label{sec:misuse_based}

Under misuse-based intrusion detection, we understand the use of a database of attack signatures to match the behavior observed under different attacks against monitored network and/or host activities in order to detect existing known attacks. A typical attack signature contains distinguishing marks of a given exploit targeting a known vulnerability. Because of the deterministic nature of this approach (i.e., a monitored activity either matches an attack signature database entry or not), misuse-based intrusion detection techniques are considered as reliable and they normally exhibit low rate of false positives. However, such techniques lack the ability to detect previously unseen, zero-day, attacks. Many existing VMM-based IDSes use a database of attack signatures to perform network intrusion detection in particular. Currently, it is a common practice to use the open-source intrusion detection engine Snort \cite{roesch:snort}, which ships with a comprehensive attack signature database that is maintained up-to-date\footnote{Snort uses the term \emph{rule} instead of \emph{signature}. As explained in the official documentation of Snort \cite{snortdoc}, the difference lies in the fact that signatures are usually constructed to detect existing attack scripts, while Snort rules aim to detect attempts to exploit existing vulnerabilities.}. Some VMM-based IDSes that use the Snort engine for network intrusion detection are proposed by Hai et al. \cite{hai:vmfence} and Roschke et al. \cite{roschke:intrusion}.  

Although misuse-based intrusion detection techniques enable reliable detection of known attacks, their application in virtualized environments suffers from a specific issue. Virtualization technology enables a single VMM to host multiple guest VMs, where each guest VM may host different OS, applications, and services, than those deployed in sibling VMs. Thus, a signature database consisting of attack signatures for all OSes, applications, and services that may be hosted on a given VMM would be of considerable size. On the one hand, a comprehensive attack signature database is needed for protection of all OSes, applications, and services that may reside at a VMM, which may result in significant delays in the attack detection as well as high overhead in terms of consumed computing and I/O resources. A delay in the detection of attacks significantly increases the possibility for intrusion. Also, the excessive consumption of resources might impair the performance of guest VMs since they share hardware resources with the IDS. In order to address these issues, a useful feature for a misuse-based VMM-based IDS is the \emph{automatic adaptation} of its attack signature database whereby attack signatures are continuously added or removed from a signature database by the IDS itself such that at each point in time the database contains signatures \emph{only} for the currently running OSes, applications, and services. 

Given the constantly changing landscape of a typical virtualized environment, it is important for an attack signature database to be continuously updated. The frequent landscape changes occur mainly due to the migration of VMs, a feature specific to virtualized environments enabling the re-deployment of VMs between VMMs. A VM may arrive at a VMM, or depart from it, at any given time resulting in the need for continuous adaptation of the attack signature database; signatures are added in case a VM has arrived, or removed in case a VM has departed. To keep their signature databases up-to-date, VMM-based IDSes normally continuously collect information about the running OSes, applications, and/or services (e.g., types, versions, and similar) employing the monitoring methods discussed in Section~\ref{sec:vmm_based}.

\begin{figure}[!t]
\centering
\includegraphics[scale=0.9]{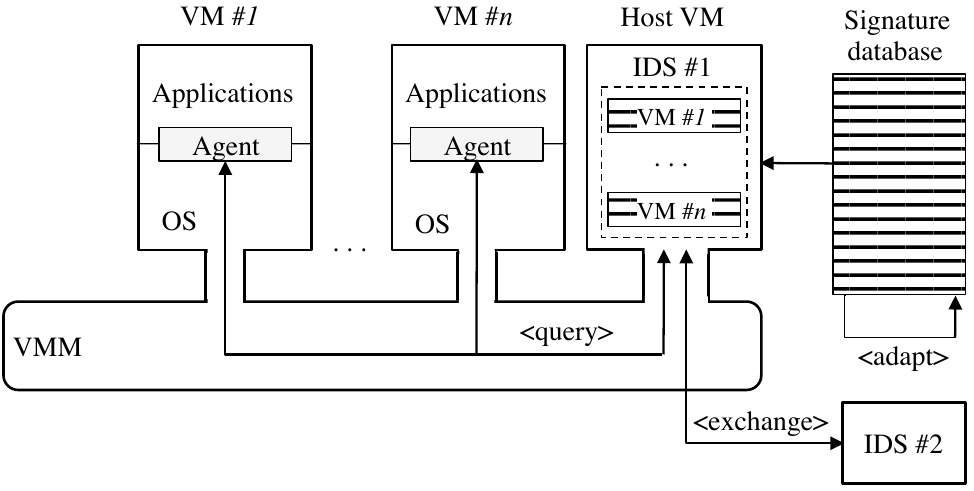}
\caption{Automatic adaptation of an attack signature database.}
\label{fig:signature_database_sizing}
\end{figure}

There are multiple approaches for automatically adapting a signature database, for example, some distributed VMM-based IDSes use protocols specifically designed to exchange signatures between IDS nodes/components residing on different VMMs. The exchanged signatures are normally related to a migrated VM; that is, one IDS node removes a set of signatures for a VM that is to be migrated, and sends them to the IDS node that resides on the VMM where the VM is to be migrated and in whose database the sent signatures must be added. For instance, VMFence \cite{hai:vmfence} uses such a signature exchange mechanism to adapt its signature database. Further, a VMM-based IDS may periodically query the guest VMs in order to renew its information about the running OSes, applications, and services so that it can accordingly adapt its attack signature database by itself (e.g., activate or deactivate a specific set of signatures with respect to the monitored environment). Note that a combination of both approaches is also possible. In Figure~\ref{fig:signature_database_sizing}, we show an example of such a case where a single node of a distributed VMM-based IDS, i.e., \emph{IDS \#1}, performs automatic signature database adaptation by periodically querying the hosted guest VMs while at the same time exchanging attack signatures with other IDS nodes such as \emph{IDS \#2}.

The automatic attack signature database adaptation is a feature which is increasingly adopted by VMM-based IDSes. The trade-offs between the advantages (e.g., shorter attack detection time) and disadvantages (e.g., introduced communication overhead due to signature exchange) over non-adaptive attack signature databases are yet to be explored in more detail.

\subsubsection{Anomaly-based Intrusion Detection}
\label{sec:anomaly_based}
\begin{figure}[!t]
\centering
\includegraphics[scale=0.9]{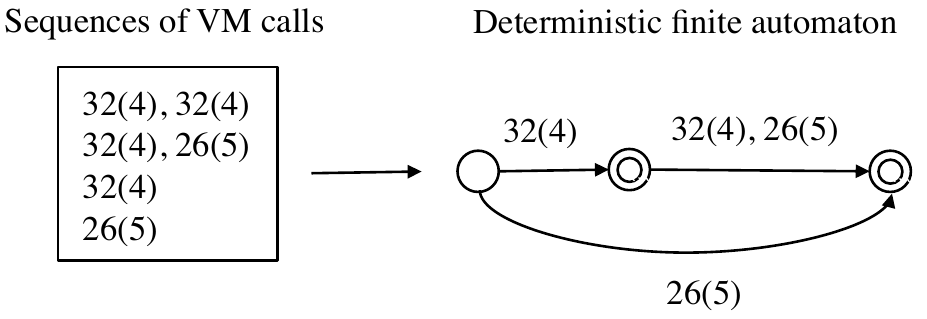}
\caption{Deterministic finite automaton for describing regular execution of the \emph{read} system call. Each VM call sequence corresponds to a benign behavior of the \emph{read} kernel service handler. [c.f. \cite{srivastava:secure}].}
\label{fig:wizard_anomaly}
\end{figure}

Anomaly-based attack detection techniques distinguish between regular and anomalous activities by comparing the observed behaviour against a reference baseline profile of ``normal'' host and/or network activities.  Thus, besides the detection of known attacks, anomaly-based techniques enable the detection of zero-day attacks, a core feature of such techniques that represents a major advantage they have over misuse-based attack detection techniques. In virtualized environments, anomaly-based attack detection techniques are normally used for host intrusion detection. For instance, the VMM-based IDS proposed by Azmandian et al. \cite{azmandian:virtual} uses clustering algorithms (i.e., the k-nearest neighbor algorithm) to classify system call sequences originating from guest VMs into regular or anomalous sequences. Some researchers, for example, Srivastava et al. \cite{srivastava:secure} and Nascimento et al. \cite{nascimento:anomaly_based}, have designed VMM-based IDSes that create models of regular system activities during an initial IDS training phase and then use them for identifying anomalous activities during operation. For instance, the Wizard IDS \cite{srivastava:secure} creates models capturing the regular execution of sequences of VM calls, where each sequence corresponds to a given system call\footnote{Wizard \cite{srivastava:secure} is a non-intrusive IDS (Section~\ref{sec:vmm_based}) which creates models consisting of VM calls intercepted at the virtualization layer.}. In Figure~\ref{fig:wizard_anomaly}, we depict a model (i.e., deterministic finite automaton) describing the sequences of VM calls that correspond to a regular execution of the \emph{read} system call (e.g., \emph{32(4), 32(4)}; \emph{26(5)}, and so on). Wizard \cite{srivastava:secure} uses this model for detection of kernel-level keyloggers that modify the regular execution of the \emph{read} system call. 

When it comes to protecting a virtualized environment, two issues unique for such an environment are of major concern: (i) ensuring security of the virtualization layer, and (ii) ensuring protection from a malicious sibling VM exploiting misconfigured or vulnerable VMM \cite{bharadwaja:axen}. Thus, the detection of attacks against VMMs is a crucial feature of VMM-based IDSes. However, documentation of attacks against VMMs and respective vulnerabilities are rare at the time of writing\footnote{We analyze attacks against VMMs in terms of their use in IDS benchmarking later in Section~\ref{sec:malicious_workloads}.}, and thus, it is a common opinion that most attacks against VMMs that occur in practice exploit undisclosed vulnerabilities. In order to detect such unknown attacks, it is a common trend for VMM-based IDSes to use anomaly-based intrusion detection techniques. Quoting from Bharadwaja et al. \cite{bharadwaja:axen} regarding the reasoning behind using anomaly-based attack detection in their VMM-based IDS: ``\textit{Our objective is to prevent intrusions via hypercalls from compromised guest VMs in order to protect the VMM and maintain uninterrupted services to other guest VMs. Since there are no documented hyper-call attacks to date, we focus the implementation of our security system on anomaly detection instead of misuse detection mechanisms.}''

Many anomaly-based attack detection techniques are already thoroughly studied in the context of intrusion detection in virtualized environments (e.g., \cite{bharadwaja:axen}, \cite{avritzer:monitoring}). Thus, in this section, we focus on a specific technique that has not yet been subject of extensive research, i.e., use of \emph{performance signatures} for attack detection in virtualized environments. This technique distinguishes between regular and anomalious system activities by monitoring for abnormal performance signatures, for example, abnormal resource consumption such as unusual high memory or CPU consumption. The use of performance signatures for attack detection in traditional environments has already been studied in the literature. For instance, motivated from earlier studies, i.e., \cite{avritzer:the_automatic} and \cite{avritzer:ensuring}, Avritzer et al. \cite{avritzer:monitoring} propose an attack detection engine that uses data provided by the Microsoft Windows Management Instrumentation API (WMI) to detect attacks against software applications. They show that performance signatures can be used to succesfully detect different classes of attacks, for example, buffer overflow, DoS (Denial-of-Service), and stack overflow attacks, in an efficient manner. 

When it comes to using performance signatures for attack detection in virtualized environments, a possible approach is to implement intrusion detection at two levels: (i) at the virtualization host layer, and (ii) at the virtual machine layer. There are multiple alternative approaches to implement the needed monitoring mechanisms in terms of the used sources of input data. An optimal alternative can be chosen with respect to the particular monitored OSes, applications, and/or services. A possible attack detection procedure is as follows: the baseline performance signatures and the associated intrusion detection logic are implemented at the virtualization host layer. Once a security intrusion is detected at this layer, additional monitoring is triggered at the VM layer to identify the VM that is the target of the detected security attack.  A set of continuous buckets (i.e., streams of monitored data) are used to ensure that each monitored performance signature is normally distributed.  The normal distribution of the performance signatures monitored at the virtualization host layer is a consequence of the law of large numbers and the superposition of the performance signatures of multiple VMs. Regarding the evaluation of the attack detection method mentioned above, an important issue is how to characterize the baseline safe environment and evaluate whether the resulting characterization is representative. 
 
\section{Requirements and Challenges for Benchmarking VMM-based IDSes}
\label{sec:challenges_on_benchmarking}

In this section, we discuss requirements and challenges related to workloads and metrics for benchmarking VMM-based IDSes. Although we already discussed some of these challenges in a previous work \cite{milenkoski:towards}, we now analyze them in greater detail, focussing on challenges that apply to non-intrusive VMM-based IDSes and some that apply to intrusive VMM-based IDSes as well. 

\subsection{Workloads}
\label{sec:workloads}

When it comes to benchmarking an IDS, one needs both malicious workloads (i.e., workloads that contain attacks) and benign workloads (i.e., workloads that contain only regular activities). One can use them separately, as \emph{pure} malicious and \emph{pure} benign workloads, or in combination as \emph{mixed} workloads. For instance, pure benign workloads can be used for evaluating the monitoring performance overhead or the capacity of an IDS as in Bharadwaja et al. \cite{bharadwaja:axen} and Hai et al. \cite{hai:avmm-based}. Pure malicious workloads can be used for evaluating the attack coverage of an IDS \cite{mell:anoverview}. Mixed workloads are normally used to subject an IDS under test to realistic attack scenarios. In this section, we investigate the requirements and challenges related to both malicious and benign workloads. Under \emph{malicious workloads}, we understand both pure malicious and mixed workloads since both of them contain attacks. Under \emph{benign workloads}, we understand pure benign workloads. 

\subsubsection{Benign Workloads}
\label{sec:benign_workloads}
\begin{figure*}[!th]
\centering
\includegraphics[scale=0.8]{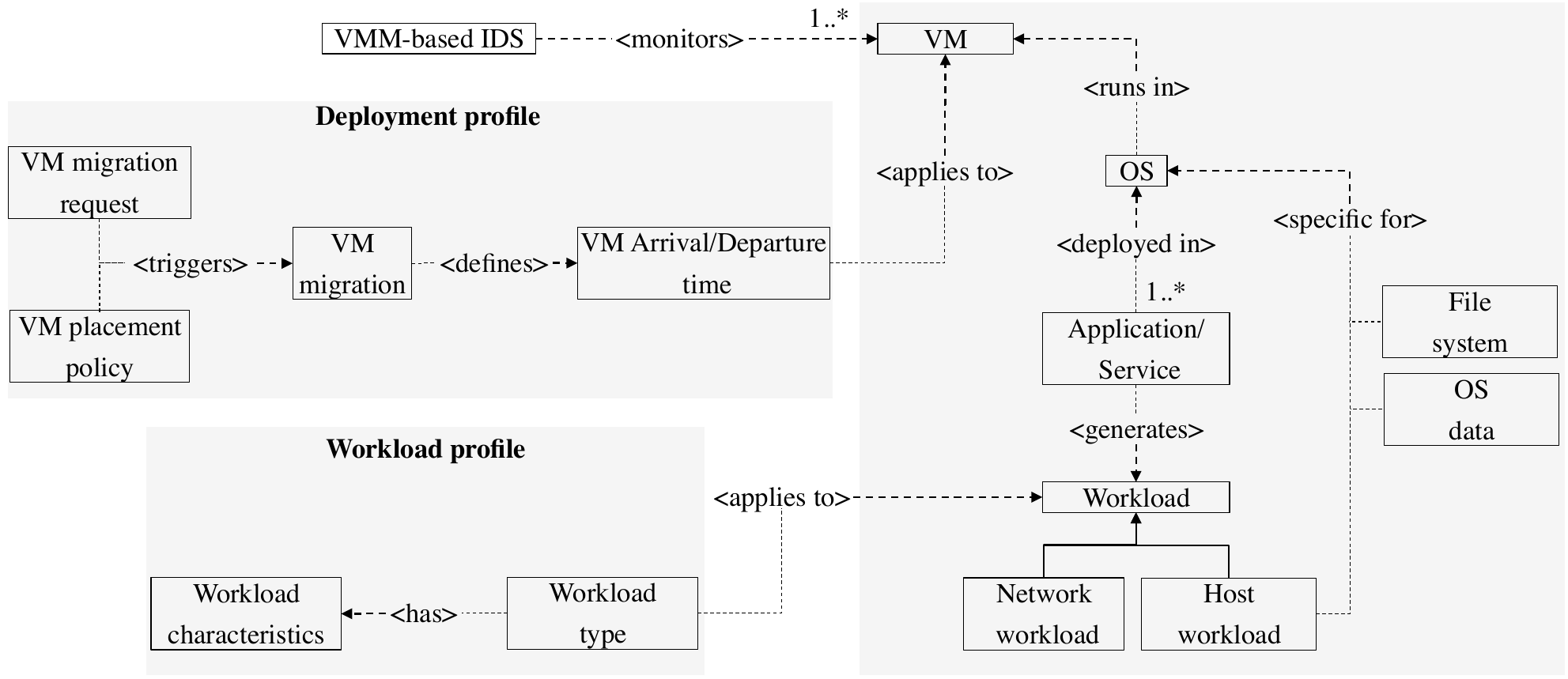}
\caption{Monitoring landscape of a VMM-based IDS.}
\label{fig:monitoring_landscape}
\end{figure*}

In order to identify the requirements and challenges related to the use of benign workloads in benchmarking VMM-based IDSes, we first analyze the typical operational environment of such IDSes. To this end, we define \emph{monitoring landscape} of a typical VMM-based IDS as a set of guest VMs hosted on a single VMM with their own separate workloads. In order to monitor the workloads of guest VMs, among many other things, a VMM-based IDS normally monitors OS-specific system components (e.g., file systems), as well as OS data stored in the main memory of the VMs (e.g., process structures, system call codes, and similar). File systems and OS data are normally monitored using host-based intrusion detection sensors (e.g., \cite{hai:vmfence}, \cite{bharadwaja:axen}). We analyze a monitoring landscape from two perspectives: (i) the type and characteristics of the workloads originating from guest VMs, and (ii) the temporal dynamicity, i.e., the changes of the monitoring landscape over time. To capture these two aspects, we define the notions of \emph{workload profile} and \emph{deployment profile} of a monitoring landscape. In Figure~\ref{fig:monitoring_landscape}, we depict the structure of the monitoring landscape of a VMM-based IDS showing the workload and deployment profiles. In the following, we discuss these profiles in the context of benchmarking VMM-based IDSes.

The applications and/or services deployed in each guest VM generate network and/or host workloads monitored and processed by the IDS. As previously mentioned (Section~\ref{sec:vmm_based}), many VMM-based IDSes have the ability to monitor both network and host workloads. The workloads originating from each guest VM are normally of a specific \emph{type} (e.g., streaming, data processing, scientific computing) with specific \emph{characteristics} relevant to intrusion detection (e.g., burstiness, throughput, and similar). The benign workload types allow for definition of realistic user activity profiles (e.g., streaming video, performing scientific calculation tasks). Among many other things, user benign activity profiles are used for developing and/or tuning benign workload generators that mimic user behavior. Because of their importance in benchmarking in general, many research works are focussing on identifying workload types that are normally seen in virtualized cloud environments. For instance, the survey performed by IBM \cite{ibm:dispelling} identifies data processing and multimedia streaming as one of the most common cloud workload types. We define the workload type and workload characteristics for each guest VM as elements of the \emph{workload profile} of the monitoring landscape. 

In a typical virtualized environment, a guest VM can arrive (or be activated) at, or depart from (or be deactivated at), a VMM at any time. We refer to these times as \emph{VM arrival time} and \emph{VM departure time}, respectively. In cloud environments, the migration of VMs is one of the main causes of changes in the number of hosted guest VMs at a VMM. Guest VMs may be migrated because of VM placement policies that aim to optimize resource efficiency during operation by automatically migrating VMs in response to changes in their workload profiles. In addition, a VM user may explicitly request VM migration. Therefore, we argue that the number of guest VMs that a VMM-based IDS monitors may change significantly over time. Given the above discussion, under \emph{depoyment profile} of a monitoring landscape, we understand the arrival and departure times of the monitored VMs as well as the VM migration features of the virtualized environment. 

In light of the above observations, it can be concluded that typical benign workloads in virtualized cloud environments as observed by a VMM-based IDS have two characteristics: (i) \emph{scalability} (i.e., benign workloads scale up or down as guest VMs arrive at, or depart from, a VMM), and (ii) \emph{heterogeneity} (i.e., the benign workloads that originate from guest VMs are a mixture of various workload types and characteristics). We argue that scalable and heterogeneous benign workloads are crucial for accurate benchmarking of VMM-based IDSes. Although we discuss the use of such workloads in IDS benchmarking in particular, their use has been studied also in other research areas such as performance benchmarking. For instance, Binning et al. \cite{binning:how} state that the use of heterogeneous benign workloads in cloud performance benchmarking is required and preffered over traditional benign workloads, which are still used in many benchmarking experiments in the field. Under traditional benign workloads, we understand workloads that are not heterogeneous and scalable in nature as previously described. 

In order to investigate the representativeness of the used benign workloads in VMM-based IDS benchmarking experiments, we surveyed multiple such experiments and observed that the benign workloads usually originate from a single VM\footnote{We surveyed in total 10 publications published in the time period of 2009-2011 describing benchmarking experiments of VMM-based IDSes. The surveyed publications are \cite{lombardi:secure, hai:avmm-based, bharadwaja:axen, hai:vmfence, riley:guest, dunlap:revirt, laureano:protecting, srivastava:secure, nascimento:anomaly_based, kong:adjointvm}.}. Further, they are generated by workload generators (e.g., SPEC CPU2000 \cite{spec:cpu2000}, iozone \cite{iozone}), or by manual task execution (e.g., use of the UNIX command \emph{dd}, kernel compilation). Such benign workloads lack scalability (i.e., they originate from a single VM) and heterogeneity (i.e., they are not a mixture of different workload types). In this section, we focus on specific benchmarking scenarios where the use of traditional benign workloads is not appropriate and scalable and heterogeneous benign workloads are needed. We also identify challenges related to the use of scalable and heterogeneous benign workloads. We investigate the use of such workloads in the following scenarios: (i) evaluation of the ability of VMM-based IDSes to handle changes in the monitoring landscape over time, and (ii) definition of a baseline ``normal'' workload profile, normally used for training anomaly-based IDSes\footnote{We do not consider this list as exhaustive. For instance, when the goal is to evaluate the attack detection accuracy of a VMM-based IDS in attack scenarios representative for virtualized environments, we consider the use of scalable and heterogeneous benign workloads as background activities mixed with the attacks.}. 

Since the number of VMs monitored by a VMM-based IDS may change over time, an important benchmarking objective is to measure the ability of the IDS to handle such changes; that is, it is important that the IDS is able to maintain operational efficiency as its monitoring landscape changes over time. A VMM-based IDS lacking such an ability may be of limited use in a real-world cloud environment. Therefore, there has been increasing research on IDSes that automatically adapt to changes in the environment. For instance, Hai et al.  designed VMFence \cite{hai:vmfence} to be able to transfer security-related information (e.g., attack detection signatures) and migrate attack detection processes between VMMs when a guest VM migration occurs. Quoting from Hai et al. \cite{hai:vmfence}: ``\textit{..., since VMFence must adapt to the
movement of VM. After the service VM is migrated successfully,
the main process should reply to such change.}'' Taking into consideration the importance of the ability of VMM-based IDSes to handle changes in the monitoring landscape, it can be concluded that scalable and heterogeneous benign workloads are required when it comes to the evaluation of this ability. For instance, one may evaluate the workload processing capacity of a VMM-based IDS in a scenario when new VMs arrive and no VMs depart. In this scenario, the capacity would be determined as the highest workload throughput that an IDS is able to handle with respect to the number of running VMs.

Benign workloads that are representative for virtualized environments may also be used for training anomaly-based VMM-based IDSes. As mentioned in Section~\ref{sec:intrusion_detection_techniques}, many existing VMM-based IDSes feature anomaly-based host intrusion detection. Anomaly-based IDSes need to be trained in order to efficiently distinguish between malicious and non-malicious activities by using both malicious and benign workloads. However, often certain relevant classes of attacks cannot be used in the training phase of an anomaly-based IDS, for example, zero-day attacks.  In such case, pure benign workloads that are considered as ``normal'' for a given environment are used for traning, whereby any deviation from such workloads is assumed as malicious. This assumption is known as the ``closed world'' assumption and it is widely used in practice. Although the ``closed world'' assumption is considered as unrealistic by some researchers (e.g., Witten et al. \cite{witten:data}), to the best of our knowledge, an alternative approach for tackling the issue mentioned above currently does not exist. 
 
 We argue that it is a significant challenge to determine what is a ``normal'' workload profile of a typical virtualized environment. Taking into account the dynamicity of the monitoring landscape, it can be concluded that the estimation of the number of VMs whose workloads are to be monitored by a VMM-based IDS at a point in time is a demanding task. Consequently, ``normal'' benign workload profiles usually cannot be constructed in a straightforward manner since many degrees of freedom need to be considered, such as VM placement policies, features of the virtualized platform (e.g., whether it allows users to explicitly request a VM migration), and similar. Further, the definition of ``normal'' benign workloads is challenging given that each VM may be used for a specific purpose (e.g., scientific computation, video streaming, and similar)\footnote{Some virtualized environments, such as the Amazon Elastic Compute Cloud (EC2) \cite{AmazonEC2}, support the creation of different types of pre-configured VMs that provide pre-packaged software stacks for a specific type of environment.
For instance, the Amazon Machine Images (AMIs) enable the creation of VMs with a deployed OS and a specific set of applications in order to accommodate a particular user activity, for example, software development.}; that is, considering arriving and departing VMs, the workloads that a VMM-based IDS has to monitor may not conform to a specific usage profile that is fixed over time. Finally, the diversity and dynamicity of the monitoring landscape poses a technical challenge. For instance, in the case when one is supposed to create a training dataset for a VMM-based IDS that monitors system calls, multiple datasets consisting of systems calls
of all OSes that may be hosted on the VMM where the IDS resides are required. Lack of such datasets would significantly impair the ability of the IDS to detect attacks against all OSes that may be deployed in the virtualized environment. 

From the above discussions one may conclude that the use of scalable and heterogeneous workloads in benchmarking VMM-based IDSes is a challenging requirement. Frameworks that generate scalable and heterogeneous benign workloads already exist, for example, at the time of writing, a prominent example is the SPECvirt\_sc2010 benchmark framework developed by SPEC (Standard Performance Evaluation Corporation) \cite{specvirt}. The SPECvirt\_sc2010 benchmark incorporates several previous SPEC benchmarks (i.e., SPECweb2005, SPECjAppServer2004, and other) to generate heterogeneous workloads. Further, to scale the workloads, SPECvirt\_sc2010 uses ``tiles'', i.e., set of VMs executing the previously mentioned benchmarks, that may be configured to be activated at a specific time. This and similar frameworks need to be evaluated in detail in terms of their applicability in benchmarking VMM-based IDSes.

\subsubsection{Malicious Workloads}
\label{sec:malicious_workloads}

In order to identify requirements and challenges when it comes to the use of malicious workloads for benchmarking VMM-based IDSes, we first analyze the characteristics of common attack scenarios against virtualized environments; we consider attacks that take place in an orchestrated sequential time order, advancing towards a common final goal. We refer to such attacks as \emph{multi-step} attacks. We define a multi-step attack as a composite attack consisting of several temporally sequential elementary attacks. Under \emph{elementary attack}, we assume an atomic, logically non-splittable attack realized through a single malicious activity, for example, use of a single attack script. 

Multi-step attacks are suitable to model attacks against virtualized environments due to the multi-layered architecture of the latter. In contrast to traditional environments, virtualized environments have an additional layer, i.e., the intermediary virtualization layer, deployed between physical hardware resources and VMs that run on top of it. A VMM is an attractive target for attackers since in case it is compromised, a VMM might provide administrative control to an attacker over all VMs hosted by the VMM. Further, an attacker may use a compromised VMM to further spread in the virtualized infrastructure of which the compromised VMM is a part. Since a VMM is normally not reachable from the ``outside world'', i.e., from outside of the virtualized platform itself, attackers usually make an attempt to intrude a given VMM by executing a multi-step attack; that is, they first intrude a guest VM hosted by the targeted VMM and then attack the target VMM itself. Ferrie \cite{ferrie:attacks} and Bharadwaja et. al \cite{bharadwaja:axen} acknowledge this attack trend and provide a brief overview of common attacks against VMMs that are normally mounted from a guest VM (e.g., VMM detection, DoS, and ``escape-to-hypervisor'' attacks). 

Beside the previously mentioned attacks, several other types of attacks are common and specific to virtualized environments. For instance, such are the attacks that exploit hardware resource sharing features of virtualized environments, for example, weak inter-VM isolation policies; an attacker may make a compromized VM to consume excessive amount of resources so that the resources available to co-located sibling VMs are significantly reduced, resulting in a DoS attack. Further, given that in cloud environments the use of allocated resources is paid according to allocation time and amount of resources, some attacks take advantage of this commercial aspect of virtualized cloud environments. For instance, in the case of a so called ``billing attack'', an attacker intrudes a guest VM in order to generate a workload that would consume high amount of resources, resulting in a bill that the authorized owner of the compromised VM would have to pay. Ferrie \cite{ferrie:attacks} gives an overview of classes of attacks that exploit features of virtualized environments. 

In the light of the above observations, one may conclude that most of the previously mentioned virtualization-specific attacks exploit operational, configurational, or design vulnerabilities in VMMs. Thus, the integrity of VMMs is crucial when it comes to securing virtualized environments. Although attacks against VMMs are currently rare\footnote{Although, at the time of writing, there are few reports of attacks against VMMs performed in practice, documented VMM vulnerabilities are not rare. The IBM's X-Force 2010 Mid-Year Trend and Risk Report \cite{ibm:midterm} documents the discovery of more than 300 vulnerabilities in VMMs. Further, the amount of practically feasible attacks against VMMs is increasing. For instance, Ding et al. \cite{ding:return_oriented} have recently shown that return-oriented programming attacks against Xen, leading to privilege escalation of a guest VM, are feasible.}, protecting against them is extremely important because of their severity. Therefore, the focus of many security researchers is shifting towards security of VMMs \cite{informationweek:vulnerability}. For instance, Szefer et al. \cite{szefer:eliminating} have worked on shrinking the attack surface of VMMs exposed to attackers, and Lombardi et al. \cite{lombardi:secure} and Bharadwaja et al. \cite{bharadwaja:axen} design VMM-based IDSes able to detect attacks targeting the VMM. Due to this trend in IDS design and development, we consider the attacks against VMMs as a requirement when it comes to the use of malicious workloads in benchmarking VMM-based IDSes.

There are many benchmarking challenges posed by the previously mentioned attack types, characteristics, and trends. For instance, the multi-step attacks normally consist of both host and network attacks and thus, many VMM-based IDSes feature network and host intrusion detection (e.g., \cite{lombardi:secure}, \cite{hai:vmfence}). In order to benchmark the attack detection accuracy of such IDSes, one needs to obtain and execute malicious workloads appropriate for both \emph{host and network} intrusion detection sensors. This is a challenging task since for each type of malicious workload, challenges of different nature apply \cite{mell:anoverview}. Further, the use of multi-step attacks as malicious workloads poses the challenge of defining an appropriate \emph{temporal order} of the attack steps with respect to the workload usage intent, i.e., which IDS property is to be measured by using such workloads. For instance, in network intrusion detection, many IDSes are constrained in their memory consumption leading to a limited capacity of network packet tracking, useful for detecting multi-step attacks. Thus, an attacker might identify IDS memory constraints and evade an IDS by delaying the execution of the planned sequential attacks and instead executing benign (non-malicious) workloads, a technique known as ``smoke-screening''. Note that the detection of composite attacks often relies on their temporal statistical characteristics \cite{wang:anovel}. Therefore, one should decide which starting times of attack steps are representative for attacks that do not factor evasion and which for attacks that do, given that \emph{attack detection efficiency} and \emph{resistance against evasion techniques} are usually considered as separate IDS benchmarking categories \cite{nsslabs:network}. In addition to the previously mentioned challenges, we focus on two others in particular: (i) using attacks against VMMs as malicious workloads, and (ii) recording and replaying workload traces that contain attacks. 

As discussed above, attacks against VMMs are important in benchmarking VMM-based IDSes due to their representativeness. It is a common approach to gather publicly available attack scripts and to then either execute them against a victim environment for live IDS testing, or to record their execution for a later replay. However, scripts for attacks against VMMs are currently rare. Although the IBM's X-Force 2010 Mid-Year Trend and Risk Report indicates the existence of more than 300 VMM vulnerabilities, it also reports that only 51 attack scripts exist at the time of writing, i.e., 2010 \cite{ibm:midterm}\footnote{Further, a substantial part of these attack scripts are attacks against virtualized gaming platforms. Such attacks are of limited use in benchmarking VMM-based IDSes, which are normally designed to protect mainly computing environments.}. Attacks against VMMs are often too complex and therefore scripts automating such attacks cannot be developed soon after the release of a security advisory. For instance, a detailed security advisory describing a new attack against Xen (CVE-2012-0217) was recently published, however, a script for this attack does not yet exist at the time of writing. Further, even when attack scripts are available, they typically require heavy modifications in order to successfully exploit the respective target environments (e.g., shellcode adjustments, return address range adjustments, and so on). Also, some attack scripts are available only for commercial purposes. For instance, a known ``escape-to-hypervisor'' attack against the VMware hypervisor is included in the attack database of the commercial penetration test tool Canvas \cite{canvas} and as such it is not publicly available. 

There are mainly two approaches that can be used to deal with the above mentioned issue of the lack of VMM attack scripts: use of a honeypot to record attacks against VMMs as executed in the real world, or use of a vulnerability injection technique to artificially inject exploitable vulnerabilities in a VMM codebase. As the use of honeypots to record malicious workloads for IDS benchmarking is a common and well established approach (e.g., as used by Asrigo et al.\cite{asrigo:using}), in this report we focus on the use of vulnerability and attack injection-based approaches, which are still in early stages of development. 
 
Vulnerability injection assumes artificial injection of exploitable software faults into the codebase of a given system. It normally consists of two major steps: first, an analysis of the target system's source code is performed so that locations where vulnerabilities can be injected are identified; then, a vulnerability is injected by performing code mutation. The vulnerability injection mechanism is normally paired with an automated attack component, i.e., an attack injector, that exploits injected vulnerabilities. Carreira et al. \cite{carreira:xception} and Rodrigues et al. \cite{rodrigues:mafalda} were the first to show that it is possible to emulate realistic hardware faults. Further research has been conducted in the domain of software fault emulation, which has recently being gaining increasing attention. For instance, Duraes et al. \cite{duraes:definition} conducted an extensive field study in order to identify types of software bugs that are usually found in software systems for the purpose of emulating software faults. Fonseca et al. \cite{fonesca:mapping} analyzed numerous security patches of web applications to discover common software faults leading to security vulnerabilities. They proposed a framework that features automatic vulnerability injection by injecting the identified faults and automated attack injection to exploit injected vulnerabilities \cite{fonesca:vulnerability}. 

Vulnerability and attack injection \cite{fonesca:vulnerability} is a promising technique for comparing IDSes as shown in \cite{fonseca:comparingids}. Vulnerability injection avoids the need for availability of vulnerable software, although there are vulnerability representativeness issues to be considered. Also, attack injection provides the emulation of malicious workloads used for IDS live testing. However, the use of these techniques in the context of VMMs is challenging and warrants further research. VMMs are considered as rudimentary OSes with a limited set of features and consequently, their codebase is considerably smaller in terms of lines of code compared to what is typical for a full-blown OS. This allows for a detailed and thorough examination of the source code of VMMs before their public release so that critical code regions that expose vulnerabilities are sanitized. 

The potential lack of locations in the code of a given VMM, suitable for injection of a vulnerable code, is often a challenging issue.  Further, even if suitable code locations are found and vulnerabilities are injected, the automatic exploitation of these vulnerabilities is also a challenging task. Since VMMs have a small codebase that is usually well examined in terms of security, they are known to be relatively free of vulnerabilities exploitable in a straightforward manner, for example, stack overflow attacks allowing arbitrary code injection and execution. Also, modern exploit preventive techniques, for example, the use of non-executable stacks, efficiently prevent such attacks against any platform, including VMMs. Thus, one may conclude that classical basic exploitation techniques are not representative for attacks against VMMs and advanced exploitation techniques that warrant extensive knowledge about the particular attacked VMM, such as code reuse techniques (e.g., return-to-libc, return-oriented programming attacks), are required. For instance, Ding et al. \cite{ding:return_oriented} show that Xen is vulnerable to return-oriented programming attacks, which require detailed knowledge on Xen's mapped address space in memory. Therefore, to ensure representativeness, attack injectors able to perform advanced exploitation are required.   

After discussing the use of attacks against VMMs as malicious workloads, we now focus on the challenge of recording and replaying workload traces that contain attacks. Among many other things, traces allow for straightforward replication of IDS benchmarking tests by trace replay without the need of repetitive execution of live attacks against a specific victim environment. The challenge of recording events relevant to IDSes in virtualized environments, i.e., VMI events, is briefly discussed by Nance et al. \cite{nance:virtual}. In this section, we analyze the challenge of recording VMI information for IDS benchmarking in more detail and provide concrete examples. 
\interfootnotelinepenalty=10000

\begin{figure}[!t]
\centering
\includegraphics[scale=0.8]{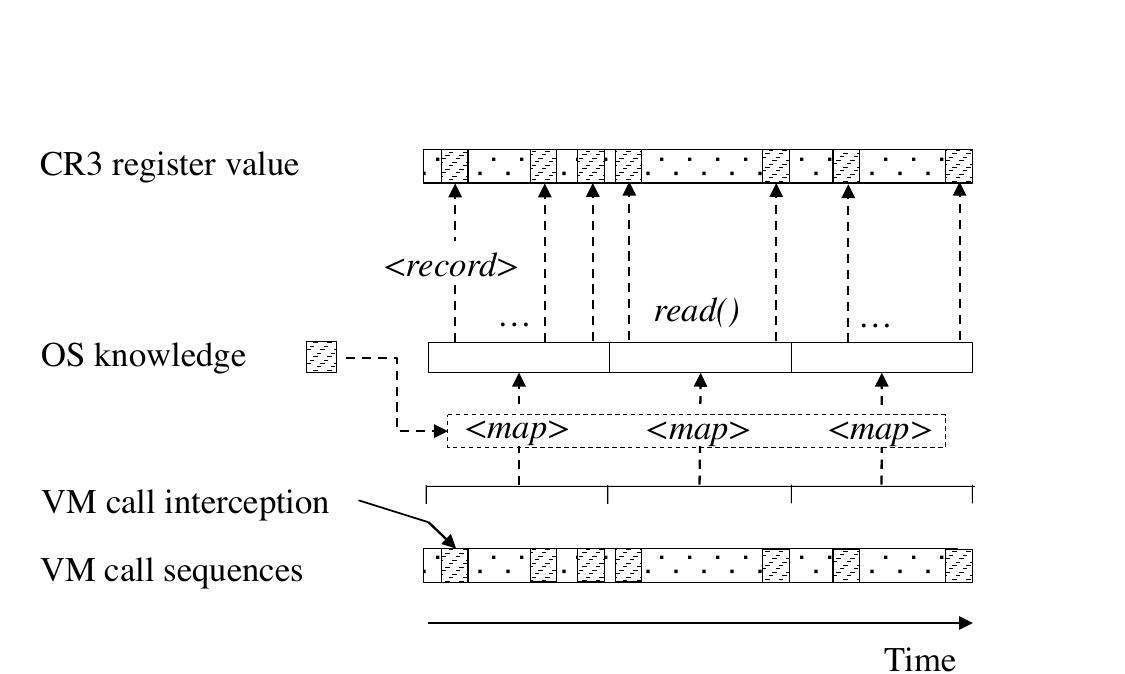}
\caption{Trace recording procedure for Wizard \cite{srivastava:secure}.}
\label{fig:trace_recording_wizard}
\end{figure}

As we stated in Section~\ref{sec:vmm_based}, non-intrusive VMM-based IDSes require both high-level context information and hardware-level information about the monitored VMs (e.g., types and versions of OSes, addresses of specific memory regions, and similar). Further, a VMM-based IDS might inspect additional information than the one it normally inspects if it suspects the existence of an ongoing attack. Also, many VMM-based IDSes, such as ACPS \cite{lombardi:secure}, feature correlation of network attacks to subsequent host attacks and issue an alert only if a correlation succeeds. We argue that the previously mentioned characteristics of VMM-based IDSes make the process of recording attacks challenging. In order to provide an illustration of this challenge, we analyze the trace recording procedure in the context of Wizard \cite{srivastava:secure}, a VMM-based IDS that we briefly described in Section~\ref{sec:vmm_based}. As a reminder, Wizard continuously monitors the execution of VM calls and for each VM call, it reads the value stored in the CR3 register to obtain information on the process that executed the call. Wizard also maps VM calls to system calls by leveraging high-level context information in order to perform intrusion detection. Thus, to record for example a trace that contains a single malicious VM call mixed with benign calls, a recording mechanism needs to record: (i) stream of VM call sequences, (ii) OS-specific high-level context information, referred to as \emph{OS knowledge}\footnote{In the context of Wizard, the OS-specific context information consists of mappings of system calls to VM call sequences obtained during the training period of Wizard  \cite{srivastava:secure}.}, and (iii) the value of the CR3 register when a VM call is intercepted. Note that the recording mechanism needs to timely capture the CR3 register value when a VM call is intercepted. In Figure~\ref{fig:trace_recording_wizard}, we depict a trace recording procedure in which the \emph{read} system call is malicious, i.e., its regular execution is modified by a kernel-level keylogger. The execution times of the VM calls that we depict in Figure~\ref{fig:trace_recording_wizard} are chosen randomly for illustration.

In the above example, we assume that a recording procedure is expected to record the exact activities required by a VMM-based IDS at a specific time. For instance, the value of the CR3 register should not be recorded when a VM call is not executed. The trace recording rate is normally constrained by many properties of the recording platform, such as the underlying I/O bandwidth of its file system. Thus, in case of highly intensive host and/or network activities, extensive logging over a longer period of time might impair the recording rate and consequently the overall quality of the generated trace files. 

Many VMM-based IDSes differ in terms of the type of information that they require. For instance, some IDSes require information on CPU registers to monitor executed system calls, while others take as input process structures stored in main memory to monitor the active processes of guest VMs. Therefore, the recording and replaying of attacks for benchmarking VMM-based IDSes is additionally complicated by the diversity in the designs of different VMM-based IDSes and the different input data that they require. Mell et al. \cite{mell:anoverview} identify a similar issue when generating malicious workloads for benchmarking traditional host IDSes.  However, since such IDSes do not require both high-level context information and hardware-level information as VMM-based IDSes do, we argue that this challenge is exacerbated when it comes to recording attacks for benchmarking VMM-based IDSes in particular. Given the complexity of the recording procedures, one may conclude that the replay procedures are equally challenging. 

\subsection{Metrics}

There are many metrics used in IDS benchmarking that quantify different IDS properties. We distinguish between two metric categories: (i) performance-related metrics, and (ii) security-related metrics. Under performance-related metrics, we understand metrics that quantify non-functional properties of an IDS under test such as capacity, performance overhead, resource consumption, and similar. Some recent IDS benchmarking experiments in which performance-related metrics are used are described in \cite{meng:adaptive}, \cite{lombardi:secure}, and \cite{mohammed:mechanism}. Since performance-related metrics have been studied extensively in the performance evaluation community, in this section, we focus on security-related metrics.

Under security-related metrics, we understand metrics that quantify properties of IDSes that are related to security concerns, such as attack coverage, attack detection accuracy, and so on. The security research community has invested significant effort on designing representative security-related metrics, which can be categorized into two groups: basic metrics and composite metrics. The basic metrics quantify individual attack detection properties such as true positive rate, true negative rate, and positive and negative predictive value. Basic metrics are often considered jointly in order to identify an optimal IDS operating point (i.e., an IDS configuration at which the value of the true positive and false positive detection rate are optimal), or to compare multiple IDSes. For example, many researchers use ROC (Receiver Operating
Characteristic) curve in order to investigate the relationship
between the true positive and false positive detection rate
of an IDS. Further, some researchers have proposed composite attack detection accuracy metrics combining the previously mentioned basic metrics. For instance, Gaffney et al. \cite{gaffney:evaluation} propose a cost-based metric and Gu et al. \cite{gu:measuring} propose a metric based on information theory.

The above mentioned performance and security-related metrics assume that the hardware resources available to an IDS under test are \emph{fixed} over time \cite{hall:capacity}. Although this assumption is valid when an IDS is deployed in a traditional environment, it does not hold when it is deployed in a modern virtualized cloud environment. One of the major advantages of cloud computing over traditional computing environments is the \emph{elastic} on-demand resource provisioning, i.e., the ability to provision and deprovision resources (e.g., computing, memory, and/or storage resources) to VMs on-the-fly according to the workload intensity and the application resource demands. As discussed in Section~\ref{sec:vmm_based}, the control and analysis components of a typical VMM-based IDS reside in the host VM, to which resources can be provisioned on-demand. For instance, the Xen virtualization platform features CPU and memory balloning, and CPU and memory hotplugging, as features enabling the on-demand resource provisioning to host and guest VMs. The CPU and memory hotplugging may be easily scripted and automatically invoked according to triggering conditions, for example, high CPU or memory utilization. Further, a VMware host VM may be configured to consume as much physical resources as it needs in order to maintain its operating efficiency\footnote{The upper bound of resources that can be used by a VMware host VM may be configured as \emph{No Hard Limit}, in which case, the host VM can theoretically consume all of the available physical resources. In practice, the resource allocation upper bound of a host VM is dynamic - it is determined by the number of hosted guest VMs and their individual configurations (e.g., the maximum amount of resources that a host VM is typically assigned does not exceed the remaining free resources after all guest VMs have been allocated a preconfigured minumum amount of resources). However, if a guest VM departs from a given VMM, the host VM can immediately use the freed resources.}.

The on-demand resource provisioning in virtualized environments diverges from the typical assumption that the amount of hardware resources used by an IDS during operation does not change over time. Therefore, in this section, we focus on analyzing this particular issue, arguing that the metrics used in benchmarking VMM-based IDSes need to take into account the elastic behavior of virtualized cloud environments. In the following, we discuss two situations showing how IDS benchmarking metrics that take into account the on-demand resource provisioning can be used to accurately quantify the capacity and attack detection accuracy of a VMM-based IDS. 

\begin{figure}
\centering
\includegraphics[scale=0.9]{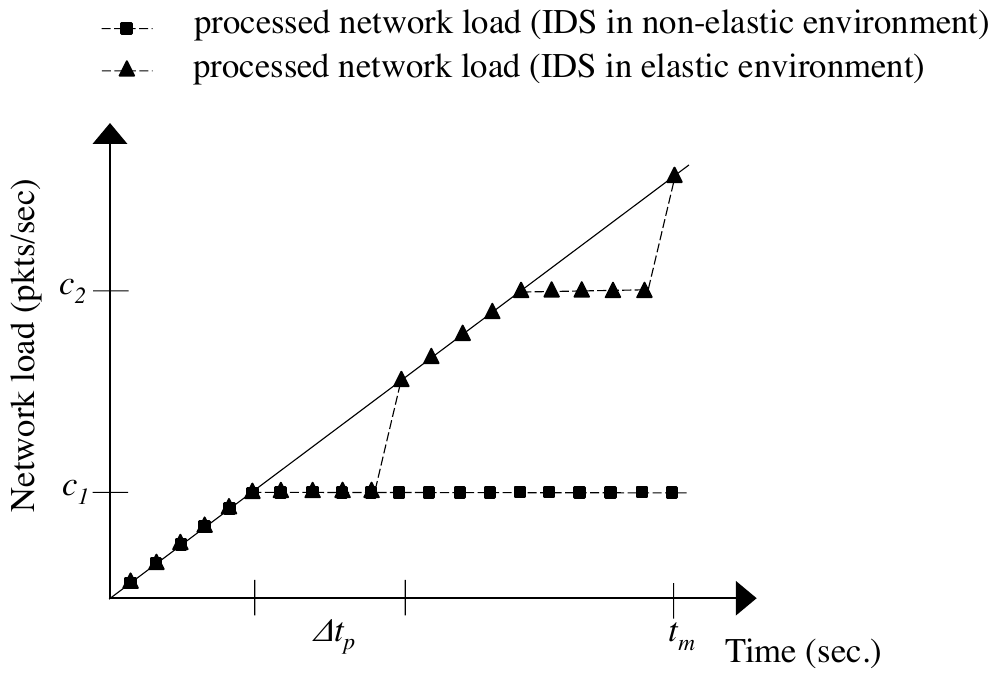}
\caption{Capacity measurements of two network-based IDSes.}
\label{fig:idsCapacities}
\end{figure}

In the first situation, we consider IDS capacity benchmarking. IDS capacity metrics usually quantify the upper bound on the performance of an IDS under increasing load assuming fixed amount of available resources \cite{hall:capacity}. During the benchmark execution, the load is constantly increased in a predefined step-wise manner until a point is reached at which the IDS starts dropping packets. An ideal elastic cloud environment would provision hardware resources to the IDS as the load increases. In some cases, that would result in only short-term degradation of the IDS performance during the time in which the new resources are provisioned. The only constraints to this continuous process are the amount of available resources that may be provisioned to the IDS and the ability of the IDS to use them. We argue that current benchmark metrics, including capacity metrics, are not always directly applicable in such situations. Some research reports on cloud performance benchmarking have also recently acknowledged similar issues \cite{binning:how}. To illustrate the previously mentioned example, we investigate the hypothetical case of benchmarking the packet processing capacities of two network-based IDSes in separate equivalent benchmark tests. This scenario is depicted in Figure~\ref{fig:idsCapacities}. Under packet processing capacity, we understand the maximal processed network load without dropping any packets.  We consider an IDS deployed in a traditional non-elastic environment and an IDS deployed in an elastic environment, i.e., a VMM-based IDS.  The network load monitored by the IDSes is expressed in packets per second (pkts/sec). It increases linearly until time $t_{m}$. We assume that the network interfaces of the systems where the IDSes run have enough bandwidth to handle the maximal network load at time $t_{m}$. Thus, the packet processing capacities of the IDSes depend on other system resources such as memory size and CPU speed.  
We assume that initially, the two IDSes reach the same packet processing saturation point $c_{1}$. In Figure~\ref{fig:idsCapacities}, one can observe that after resource provisioning in time $\Delta t_{p}$, the IDS running in the elastic environment can again process all packets until it reaches another saturation point $c_{2}$. In contrast, the IDS in the traditional environment continues to drop packets. Thus, its packet processing capacity can be unambiguously determined as $c_{1}$, which does not hold for the VMM-based IDS.  

In the scenario presented above, one may consider to use a virtualized environment without elastic properties. We argue that such an environment is not appropriate for benchmarking VMM-based IDS for cloud environments. 
Potential IDS users would compare the capacity metric values with the profile of the IDS deployment environment. As a result, they would make a decision whether the IDS satisfies the requirements to efficiently operate in that environment. 
In reality, cloud platforms are elastic. Further, each cloud environment has different elasticity properties. Thus, one cannot expect that an IDS deployed in an elastic environment would conform to the measurements taken in a non-elastic environment. 

In the second situation, we consider measuring the attack detection accuracy of a VMM-based IDS\footnote{\emph{Attack detection accuracy} is to be distinguished from \emph{attack coverage}. Under attack coverage, we understand the attack detection rate of an IDS without background benign activity. Under attack detection accuracy, we understand the attack detection rate of an IDS under usual working conditions, i.e., when attacks are mixed with background benign activities \cite{mell:anoverview}.}. We assume that the VMM-based IDS under test has adaptive characteristics, i.e., it is able to adapt its configuration and/or operation with respect to the type and amount of resources that become available to it at run-time in order to achieve optimal performance. In such case, the measured attack detection accuracy of the IDS under test would be relative to the varying amount of used resources for intrusion detection during the test run. Note that many IDSes, also traditional ones, have adaptive characteristics similar to the one mentioned above. For instance, Ragsdale et al. \cite{ragsdale:adaptation} propose an IDS that invokes new attack detection engines in case the amount of resources available to it increases during operation. Further, even an IDS that is not designed with adaptability in mind may exhibit improved performance in case it is provisioned with additional resources at run-time. For instance, if additional memory is provisioned to the host VM, the deployed IDS engine, being a host VM process, would have more available memory for use. Thus, time-critical operations, such as buffering network packet fragments, might be performed with greater speed resulting in detection of many attacks that can be missed in case of limited available main memory, especially when the IDS monitors high volume network activities. 

In light of the above observations, we argue that to quantify properties of a VMM-based IDS, one should use metrics that explicitly take into account the elasticity characteristics of the deployment environment, i.e., the type and amount of provisioned resources to the IDS under test during operation. Such metrics would also be able to quantify the difference in the measured IDS properties when the IDS under test is deployed in virtualized environments with different elasticity features. However, the construction of representative elasticity metrics is a challenging issue, currently being of extensive research in the performance evaluation community. A number of issues are actively debated, such as determining a proper view of the elastic system under test (e.g., black-box vs. white-box), determining which specific system features and characteristics elasticity metrics should reflect (e.g., scalability, precision/speed of resource provisioning), and so on.

\section{Conclusion}
\label{sec:conclusion}

In this work, we surveyed the state-of-the-art on IDSes designed specifically to operate in virtualized environments (VMM-based IDSes), focussing on intrusive and non-intrusive VMM-based IDS architectures. Further, we analyzed misuse-based and anomaly-based intrusion detection, the two most common intrusion detection techniques, in terms of how the existing VMM-based IDSes apply them in order to detect attacks. In this direction, we discussed innovative features of VMM-based IDSes, such as the automatic attack signature database adaptation and the use of performance signatures for attack detection. 

After the analysis of the current trends in intrusion detection in virtualized environments, we identified specific requirements in benchmarking VMM-based IDSes. This includes the use of attacks against VMMs as malicious workloads, IDS benchmarking metrics that explicitly take into account elasticity aspects, the use of scalable and heterogeneous benign workloads, and so on. Also, we identified challenges related to the previously mentioned requirements such as the definition of elasticity metrics for use in IDS benchmarking, obtaining attacks against VMMs in an executional or trace form, and the definition of ``normal'' usage profiles of virtualized environments. These challenges may be used as pointers to future research efforts contributing towards accurate and efficient benchmarking of VMM-based IDSes. As part of our work, we plan to provide concrete solutions for the presented challenges and to study promising intrusion detection techniques and methods for generation of malicious workloads, for example, performance attack signatues, vulnerability and attack injection in VMMs, and so on.

\cleardoublepage
\renewcommand\bibname{References}
\addcontentsline{toc}{section}{\bibname}

\bibliographystyle{plain}
\bibliography{SPEC-RG-2013-002-BenchmarkingVMMBIDSes}

\end{document}